\documentclass[a4paper,twocolumn]{article}

\title{On the VC-dimension of neural networks with binary weights}

\author{
  S.~Mertens\thanks{\small email: stephan.mertens@physik.uni-magdeburg.de}\mbox{\ } and
  A.~Engel\thanks{\small email: andreas.engel@physik.uni-magdeburg.de}\\[1ex]
  {\small\it Institut f\"ur Theoretische Physik, Otto-von-Guericke Universit\"at}\\
  {\small\it Postfach 4120, D-39016 Magdeburg, Germany}
}
                         
\usepackage{graphicx,fancyheadings,float}
\usepackage[centertags]{amstex}
\usepackage[normalem]{ulem}
\date{August 1996}

\floatstyle{ruled}
\restylefloat{figure}
\restylefloat{table}

\newcommand{\dvc}{d_{\mathrm{VC}}}
\newcommand{\dct}{\dvc^{\mathrm{ct}}}
\newcommand{\dpt}{\dvc^{\mathrm{pt}}}
\newcommand{\dvct}{d_{\mathrm{VC}}^{\mathrm{typ}}}
\newcommand{\Deltatyp}{\Delta^{\mathrm{typ}}}
\newcommand{\alphavc}{\alpha_{\mathrm{VC}}}
\newcommand{\alphavctyp}{\alphavc^{\mathrm{typ}}}
\newcommand{\sign}{\operatorname{sign}}
\newcommand{\pat}[1]{\texttt{\small #1}}

\begin{document}

\maketitle

{\bf Abstract: }{\it
We investigate the VC-dimension of the perceptron and simple
two-layer networks like the committee- and the parity-machine
with weights restricted to values $\pm1$.
For binary
inputs, the VC-dimension is determined by atypical pattern sets, i.e.\
it cannot be found by replica analysis or numerical Monte Carlo sampling.
For small systems, exhaustive enumerations yield exact results. For systems
that are too large for enumerations, number theoretic arguments give lower
bounds for the VC-dimension. For the Ising perceptron, the VC-dimension
is probably larger than $N/2$.
}

\section{Introduction}
\label{sec:intro}

Presently investigations in different fields including mathematical
statistics, computer science and statistical mechanics aim at a
deeper understanding of information processing in artificial neural networks. 
Every field has developed its own concepts which although related to each other
are naturally not identical. In order to use (and appreciate) progress made
in another field it is hence important to know the different concepts and
there mutual relation. 
The Vapnik--Chervonenkis--(VC--) dimension is one of the central quantities
used in both mathematical statistics and computer science 
to characterize the performance of classifier systems \cite{vapnik:chervonenkis:71,vapnik:82}.
In the case of feed-forward neural networks it establishes connections
between the storage and
generalization abilities of these systems \cite{haussler:etal:91,parrondo:vdbroeck:93,engel:94}.
Unfortunately, for most
architectures the precise value of the VC-dimension ist not known and only bounds
exist \cite{baum:haussler:89}.

The VC--dimension was introduced to characterize certain {\em extreme} situations
in machine learning. It is therefore very useful to derive bounds for the
network performance by considering the worst possible case. 
Complementary investigations in statistical mechanics focus on the {\em typical}
behaviour described by appropriate averages. In simple situations as
provided, e.g, by the spherical perceptron it turns out that the typical and
worst case behaviour are not dramatically different \cite{engel:vdbroeck:93}. It is then 
comparatively easy to establish connections between results obtained in
different fields.

In the present paper we discuss some peculiarities that are encountered when 
analysing the VC-dimension of neural networks with binary weights. Binary
weights are the extreme case of discrete couplings with obvious advantages
in biological and technical implementations. It turns out, however, that in
this case 
the typical and the extreme behaviour of the network can be rather different.
Therefore  the relation between results obtained by different approaches is
less obvious.

Let us also note that in the mathematical literature binary weights are usually assumed
to take on the values $0$ and $1$. Physically minded people on the other hand prefer the
values $-1$ and $1$ reminescent of spin systems. As we will show, these two choices
make a difference.

The paper is organized as follows: After giving the basic definitions in the
next section we discuss some simple examples in section \ref{sec:examples}.
In section \ref{sec:analytics} we give a short
discussion of analytical methods using the replica trick to
calculate the behaviour of the typical growth function.
Section \ref{sec:typical} is devoted to the 
numerical investigation the typical growth function for the
binary perceptron and simple two-layer networks.
In section \ref{sec:exact} we
derive bounds for the VC--dimension of neural
networks with binary couplings including simple multilayer systems. 
The bounds show that the VC-dimension is determined by \emph{atypical} situations.
The VC-dimension can hence not be inferred
from the properties of the typical growth function. 
We give arguments that the value of the VC-dimension for networks with binary weights
may depend on whether the
input vectors are continuous, binary $(0,1)$ or binary $(-1,1)$. 
Finally section \ref{sec:coda} contains our conclusions.

\section{Basic definitions}
\label{sec:basics}

The VC-dimension $\dvc$ is defined via the growth function $\Delta(p)$.
Consider a set $X$ of instances $x$  and a set $C$ of (binary)
classifications $c:x \to \{-1,1\}$ that group all $x \in X$ into two classes
labeled by $1$ and $-1$ respectively.
In the case of feed-forward neural networks \cite{hertz:etal:91} with $N$ input units and one
output unit $X$ is the space of all possible input vectors $\boldsymbol{\xi}\in\Bbb{R}^N$ or
$\boldsymbol{\xi}\in\{-1,+1\}^N$, the
class is defined by the binary output $\sigma = {\pm 1}$ and $C$ comprises all
mappings that can be realized by different choices of the couplings $\mathbf{J}$
and thresholds $\boldsymbol{\theta}$ of the network.
For any set $\{x^{\mu}\}$
of $p$ different inputs $x^1,\ldots, x^p$ we determine the number
$\Delta ( x^1, \ldots, x^p)$ of different output vectors 
$\{\sigma_1,\ldots, \sigma_p\}$
that can be induced by using all the possible classifications $ c\in C$.
A pattern set is called {\em shattered} by the class $C$ of classifications
if $\Delta ( x^1, \ldots, x^p)$ equals $2^p$, the maximal possible number of
different binary classifications of $p$ inputs. 
Large values of $\Delta (x^1, \ldots, x^p)$ hence roughly correspond
to a large diversity of mappings contained in the class $C$. The growth
function $\Delta (p)$ is now defined by
\begin{equation}\label{Delta}
\Delta (p) = \max_{\{x^\mu\}} \Delta (x^1,\ldots, x^p)
\end{equation}

It is clear that $\Delta (p)$ cannot decrease with $p$. Moreover for small $p$
one expects that there is at least one shattered set of size $p$ and hence 
$\Delta (p) = {2^p}$. On the other hand this exponential increase of the growth function
is unlikely to continue for all $p$. The value of $p$ where it starts
to slow down should give a hint on the complexity of the class $C$ of binary
classifications. In fact the Sauer lemma \cite{vapnik:chervonenkis:71,sauer:72} 
states that for all
classes $C$ of binary classifications there exists a natural number $\dvc$
(which may be infinite) such that
\begin{equation}\label{saulem}
  \Delta(p)\left\{\begin{array}{lcl}
             =2^p &\qquad \mbox{if}\qquad & p\leq \dvc\\
             \leq \sum_{i=0}^{\dvc}{p \choose i}&\qquad
             \mbox{if}\qquad & p\geq \dvc \end{array}  \right.
\end{equation}
$\dvc$ is called the VC-dimension of class $C$. Note that it will in
general depend on the set $X$ of instances to be classified. Hence in the
case of neural networks there can be different values of $\dvc$ for the same
class of networks depending on whether the input patterns are real or binary
vectors. 

Due to the max in eq.(\ref{Delta}) it is possible that the VC-dimension is
determined by a
single very special pattern set. In many situations emphasis is, however, on
the {\em typical} properties of the system. In order to characterize the
typical storage and generalization abilities of a neural
network a probability measure $\cal P$ on the input set $X$ is introduced.
One then asks for the properties of the {\em typical} growth function
$\Deltatyp(p)$ which
at variance with eq.(\ref{Delta}) is defined as the most probable value of
$\Delta (x^1,\ldots, x^p)$  with respect to the measure $\cal P$.
In the relevant limit of large dimension $N$ of
the input space it is generally assumed that the distribution of
$\Delta(x^1,\ldots, x^p)$ is sharply peaked around this value.
In the same limit $N\to\infty$ methods from statistical mechanics can be used
to investigate the properties of $\Deltatyp(p)$.
This limit is non-trivial if $\alpha = p/N =0(1)$
and results in $\dvc = 0(N)$. We will call $\alphavc = \lim_{N\to\infty}\dvc/N$ the
VC-capacity of the neural network \cite{opper:95}.
In addition we may define $\dvct$ as the value of o $p$ at which
$\Deltatyp(p)$ starts to deviate from $2^p$ and
$\alphavctyp=\lim_{N\to\infty} \dvct /N$. The storage threshold $p_c$
is as usual defined by $\Deltatyp(p_c)/2^{p_c} =1/2$ and 
$\alpha_c=\lim_{N\to\infty} p_c /N$ is the storage capacity.

Using  Stirlings formula in eq.(\ref{saulem}) and replacing the
sum by an integral one can show that for large $N$ the relative deviation of
the upper bound from $2^p$ 
becomes $O(1)$ if 
$\alpha>2\alphavc$ (see section \ref{sec:analytics}). Since we always have
$\Deltatyp(\alpha)\leq\Delta(\alpha)$ this implies 
\begin{equation}\label{alinequ}
\alpha_c\leq 2\alphavc
\end{equation}

In this paper we concentrate on three sets of classifiers: the Ising percpetron,
the Ising committee tree and the Ising parity tree. The \emph{Ising perceptron} realizes
the classification $\boldsymbol{\xi}\mapsto\pm1$ via
\begin{equation}
  \label{eq:ising_perc}
  \sigma = \sign(\sum_{i=1}^NJ_i\xi_i-\theta)
\end{equation}
with weight vector $\mathbf{J}\in\{\pm1\}^N$. The perceptron is a prototype of what is usually
termed single layer feed forward networks: The $N$ input values $\xi_i$ are summed up and the resulting
``field'' is passed through a nonlinear function to yield a single output value $\sigma$.
The computational capabilities of single layer feed forward networks are rather limited.
Hence one is interested in multi-layer networks, where the output of single layer networks
is used as input for another single layer network. The \emph{Ising committee-machine} and the
\emph{Ising parity-machine} are examples for 2-layer networks. In both machines, the input values
$\xi_i$ are mapped to $K$ binary values $\tau_k$ by $K$ Ising perceptrons. The $\tau_k$ are called
internal representation of the input. The internal representation is mapped onto the final output $\sigma=\pm1$
by the so called decoder function in the output-layer.
The decoder function is different in both machines: The committee-machine uses
a perceptron with all weights $+1$,
\begin{equation}
  \label{eq:committee}
  \begin{split}
  \sigma &= \sign(\sum_{k=1}^K\tau_k)\\
         &= \sign(\sum_{k=1}^K\sign(\sum_{i=1}^NJ_i^{(k)}\xi_i-\theta))
  \end{split}
\end{equation}
where $\mathbf{J}^{(n)}$ is the weight vector of the perceptron that ``feeds'' the $k$-th
hidden node. The restriction to all weights $+1$ in the output perceptron is not as severe
as it may appear: The storage properties of this architecture are the same as for a machine where
the output perceptron is an arbitrary Ising perceptron (see appendix \ref{sec:symmetry}).

The parity machine simply takes the parity of the internal representation:
\begin{equation}
  \label{eq:parity}
  \begin{split}
  \sigma &= \prod_{k=1}^K\tau_k \\
         &= \prod_{k=1}^K\sign(\sum_{i=1}^NJ_i^{(k)}\xi_i-\theta)
  \end{split}
\end{equation}

\begin{figure}[htb]
  \includegraphics[width=\columnwidth]{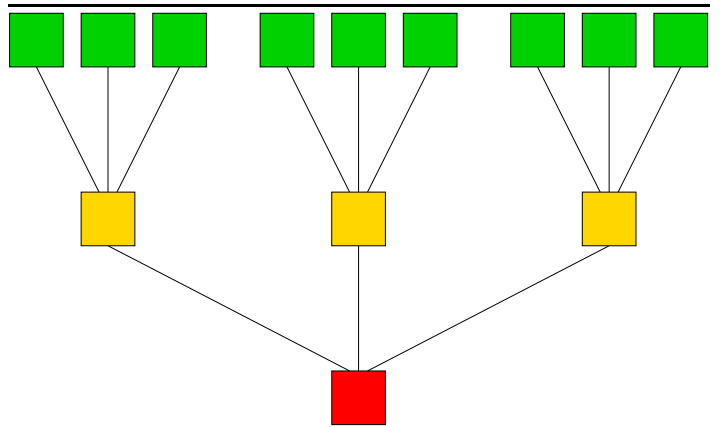}
        \caption[Fig1]{\label{fig:tree}Feedforward 2-layer network with tree structure and 
        9 input nodes, 3 hidden nodes
        and 1 output node. }
\end{figure}

In general, a hidden node can receive input from all input nodes. In this case we have
$NK$ weights to specify. If the input nodes are distributed among the hidden nodes 
such that no input node feeds more than one hidden node, the net has a tree structure
(see fig.~\ref{fig:tree}). For simplicity we will assume that the input nodes are
distributed evenly among the hidden nodes, i.e.\ each subperceptron has $N/K$ weights.

\section{Some simple examples}
\label{sec:examples}

To begin with let us discuss some simple examples.
In the case of the spherical perceptron defined by eq.\~(\ref{eq:ising_perc})
but now with
$\mathbf{J} \in \Bbb{R}^{N}$, $\sum_{j} J_j^{2} = N$ the
exact results $\dvc = N + 1$ and $\alpha_c=2$ have been obtained analytically
\cite{cover:65}. Moreover it is well known that the
number of different realizable output vectors (dichotomies) is the same for
all input pattern sets in general position \cite{cover:65}. Hence the max in
eq.(\ref{Delta}) is realized by {\em almost all} possible inputs sets of length $p$ and
$\Deltatyp(\alpha)=\Delta(\alpha)$. Furthermore (\ref{alinequ}) is
satisfied as equality. 

A particular simple pattern set for which the result for the 
VC--dimension can easily be verified is given by
\begin{equation}\label{patternset}
\begin{split}
\boldsymbol{\xi}^0 &=(0,0,0,\ldots,0)\\
\boldsymbol{\xi}^1 &=(1,0,0,\ldots,0)\\
\boldsymbol{\xi}^2 &=(0,1,0,\ldots,0)\\
\vdots\\
\boldsymbol{\xi}^N &=(0,0,\ldots,0,1).
\end{split}
\end{equation}
An arbitrary output vector $(\sigma_0, \sigma_1, \ldots, \sigma_N)$ can 
be realized for these inputs by using $J_j = \sigma_j$ and 
$\theta =-\sigma_0/2$.

Another interesting example is provided by a perceptron (\ref{eq:ising_perc})  
for which
the couplings are constrained to take the values $J_j = \{0,1\}$ only. Using
the set of input patterns described above but omitting $\boldsymbol{\xi}^0$
an arbitrary output string $(\sigma_1, \ldots, \sigma_N)$ can be
realized by using $J_j =(1+\sigma_j)/2$ and $\theta = - 1/2$.
Therefore $N\leq\dvc\leq N+1$. On the other hand it is
known that the storage capacity of this perceptron 
is given by $\alpha_c = 0.59$ \cite{gutfreund:stein:90}. This large difference between
$\alpha_c$ and $2 \alphavc$ is due to the fact that the VC-dimension is
determined by a very special pattern set and that 
$\Deltatyp(p)$ is much smaller than $\Delta(p)$. Hence the number of 
realizable output vectors
is no longer the same for all input vectors in general position.

Finally we consider the so-called Ising-perceptron, again described by 
eq.(\ref{eq:ising_perc}) but now with the constraint $J_j = \pm 1$ on the couplings. 
Since the couplings used above to show that the pattern set
(\ref{patternset}) is shattered by a spherical perceptron fulfill
this constraint it is clear that the VC-dimension of the Ising-perceptron 
is for patterns $\boldsymbol {\xi} \in \Bbb{R}^{N}$ equal to $N + 1$, exactly as for the
spherical perceptron. For $\theta = 0$ we get $\dvc = N$ in both cases. 

For binary input patterns $\xi_{i} = \pm 1$ we transform the pattern set (\ref{patternset})
according to $\xi_i \to 2\xi_i-1$. Every output vector 
$(\sigma_0, \sigma_1, \ldots, \sigma_N)$ can then be
realized by using $J_j=\sigma_j$ for $j=1,...,N$ and 
$\theta=-\sigma_0-\sum_j\sigma_j$. Therefore the VC--dimension 
is again $\dvc=N+1$. 
However, since much of the interest in neural networks with discrete 
weights is due to their easy technical implementation it is not 
consistent to design an Ising-perceptron with a threshold of order $N$.
More interesting is the determination of the VC-dimension of the
Ising-perceptron without (for $N$ odd) or with a {\em binary} threshold 
$\theta=\pm 1$ (for $N$ even) for binary
patterns. This is a hard problem (see section \ref{sec:exact}). 

We note that 
the storage capacity of the Ising perceptron has been shown to 
be $\alpha_c = 0.83$ \cite{krauth:mezard:89}. Hence also in this case we have
$\alpha_c<2\alphavc$ and the VC--dimension is not determined by
typical pattern sets. We also note that the storage capacity is believed 
to be the same for binary and Gaussian patterns
 \cite{krauth:mezard:89,krauth:opper:89,derrida:etal:91}. 
As we will see in section \ref{sec:exact}, it is unlikely that this holds also for the VC--dimension.

\section{Analytical methods}
\label{sec:analytics}

Let us fix a particular set $\{\boldsymbol{\xi}^1, \ldots,  \boldsymbol{\xi}^p \}$ of input patterns fed 
into a neural network with parameters $\mathbf{J}$. Different values of the parameters
will result in different output strings $\{\sigma^1, \ldots, \sigma^p \}$ and 
hence the input patterns induce a partition of the parameter space into
different cells labeled by the realized output sequences $\{\sigma^\mu\}$.
The cells have a certain volume $V(\{\sigma^\mu\})$ which might be
zero if the output string $\{\sigma^\mu\}$ cannot be realized. An interesting
quantity is the number of cells of a given size
\begin{equation}
{\cal N}(V) = \mbox{Tr}_{\{\sigma^\mu\}} \delta(V-V(\{\sigma^\mu\}))
\end{equation}
which, of course, still depends on the particular set of input patterns
$\{\boldsymbol{\xi}^\mu\}$. It is possible to calculate the typical value of ${\cal N}(V)$ for
randomly chosen $\{\boldsymbol{\xi}^\mu\}$ using multifractal methods and an interesting
variant of the replica trick \cite{monasson:okane:94}. This calculation has been
explicitly performed for both the spherical and the Ising perceptron
\cite{engel:weigt:96,weigt:engel:96} and we give in this section a brief summary of the results
relevant for the present paper.

For the perceptron (\ref{eq:ising_perc}) (with $\theta = 0$ for simplicity) we have
\begin{equation}
V (\{\sigma^\mu\}) = {\int d \mu (J) \prod_{\mu} \theta (\sigma^\mu \mathbf{J}
\boldsymbol{\xi}^\mu)}
\end{equation}
where $\int d \mu (J) = (2\pi e)^{-N/2} \int \prod_j dJ_j\; \delta({\sum_j}{J^2_j} - N)$
for the spherical perceptron and $\int d \mu (J) =2^{-N}{\sum_{J_j = \pm 1}}$
for the Ising case. The natural  scale of $V$ for $N\to\infty$ is then
$2^{-N}$ and it is convenient 
to introduce $k (\{\sigma^\mu\}) =-1/N \log_2 V (\{\sigma^\mu\})$ as a
measure for the size of the cells. Similarily the number of cells is
exponential in $N$ and we therefore use
\begin{equation}
c (k) = \frac{1}{N} \log_2 {\cal N} (k) = \frac{1}{N}\log_2
              \mbox{Tr}_{\{\sigma^\mu\}} \delta(k-k(\{\sigma^\mu\}))
\end{equation}
to characterize the cell size distribution. 
Realizing that $c(k)$ is the microcanonical entropy of
the spin system $\{\sigma^\mu\}$ with hamiltonian $Nk(\{\sigma^\mu\})$ it can be
calculated from the free energy
\begin{equation}\label{f}
f (\beta) = - \frac{1}{\beta N} \log_2 \mbox{Tr}_{\{\sigma^\mu\}} 
2^{-\beta Nk(\{\sigma^\mu\})}
\end{equation}
via Legendre-transform
\begin{equation}
c (k) = \mbox{min}_{\beta} [ \beta k - \beta f (\beta)].
\end{equation}
From the experience with related systems (\cite {mezard:etal:87})
one expects $f$ (and therefore $c$) to be self-averaging with respect to the
distribution of the input patterns ${\boldsymbol{\xi}^\mu}$. The average of $f (\beta)$
over the inputs can be performed using the replica trick. Within a special
replica symmetric ansatz the calculation of $ f(\beta)$ can be reduced to a
saddle-ponit integral over one (for the spherical) or two (for the Ising
case) order parameters, which are evaluated numerically \cite{engel:weigt:96}.

\begin{figure}[htb]
  \includegraphics[width=\columnwidth]{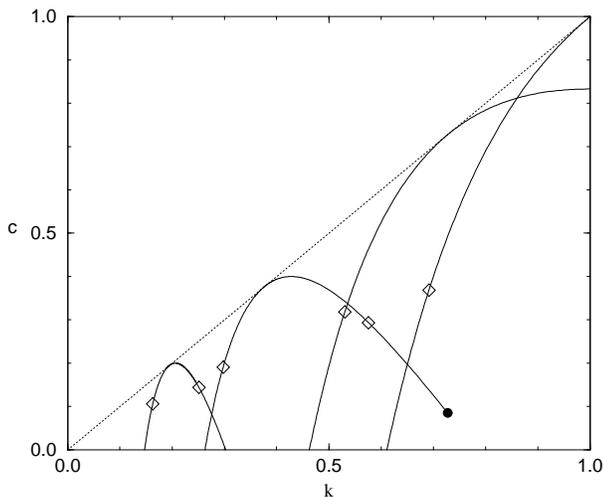}
        \caption[Fig1]{\label{fig:cells}
Distribution of cell sizes $c(k)$ in the 
coupling space of an Ising perceptron with loading ratios 
$\alpha=0.2,0.4,0.833,1.245$ (from left to right). Inside the region
given by the diamonds replica symmetry holds. The dot marks the
divergences of negative moments.
}
\end{figure}

Figure \ref{fig:cells} shows some of the resulting curves for the Ising case. For 
$\alpha = 0.2$ and $0.4$ the corresponding curves for the spherical perceptron are
rather similar. The typical cell size is given by $k_0 =$argmax $c(k)$. Therefore
$V_0=2^{-Nk_0}$ coincides with the typical phase space volume as calculated by a
standard Gardner approach \cite{gardner:88}. On the other hand $2^{Nc(k_0)}$
gives up to exponentially small countributions from other cell sizes the
typical total number $\Delta^{typ}$ of cells as determined for the spherical perceptron by
Cover \cite{cover:65}. From the explicit formulae one can show that for the spherical
perceptron $c (k_0) = \alpha $ as long as $k_0 < \infty$, i.e.
$V_0>0$, and $c (k) < \alpha $ if $k_0 = \infty$, i.e. $V_0=0$. 

For the Ising perceptron there is a smallest possible cell size
$k_{{\mathrm max}}=1$ where only one coupling remains. Hence
$\Delta^{typ} \sim 2^{Nc(k_0)}$ if $k_0 \leq 1$ and
$\Delta^{typ} \sim 2^{Nc(1)}$ if $k_0 > 1$.
The borderline $k_0 = 1$ is realized for $\alpha = 0.83$ the well known
value of $\alpha_c $ \cite{krauth:mezard:89}.
The calculation of the curves $c(k)$ therefore establishes the connection
between the two complementary approaches by Cover and Gardner to determine
the storage capacity of neural nets. 

Since one has direct access to the
number of realizable output sequences it is tempting to use this approach
also to calculate the VC-dimension
analytically. Due to the averages over the input distribution necessary to
accomplish the calculation  we can at most hope to determine
${\alphavctyp}$ in this way. As discussed above 
${\alphavctyp}$ will only coincide with $\alphavc$ if the maximum in
eq. (\ref {Delta}) is realized by a typical set of input patterns. To
determine $\alphavctyp$ we have to find the value of $\alpha$ at which
the total number of cells $\Delta^{typ}$ starts to deviate from $2^{\alpha N}$. For
$\alpha_{VC} < \alpha < 2\alpha_{VC}$ an asymptotic analysis of the bound
in eq.(\ref{saulem}) reveals that \cite{hertz:etal:91}
\begin{equation}
\frac {2^{\alpha N} - {\sum^{\alpha^{VC} N}}_{i = 0} {\alpha N \choose i}}
         {2^{\alpha N}} \to \frac{1}{2} \mbox{erfc} 
                                    \left(\sqrt{\frac{\alpha N}{2}}(\frac{2}{\alpha}-1)\right )
\end{equation}
Hence it may happen that the relative deviation is exponentially small in $N$.  
In principle we
are able to detect this deviation by using the {\em whole function} $c(k)$.
However, for very small and very large $k$ the calculation  of $c(k)$
necessitates replica symmetry breaking (\cite{engel:weigt:96}) which renders the
calculation practically impossible.

But there is another way to get some information on $\alphavctyp$ from the
$c(k)$ curves.
It is clear from eq.(\ref{f}) that $f(\beta)$ will diverge for all $\beta< 0$ 
if some of the cells are empty, i.e.\ if $k(\{\sigma^\mu\})=-\infty$. 
For $\alpha < \alphavctyp$ this is
possible only if the patterns are linearily dependent. For Gaussian patterns
the probability for this to happen is zero and therefore no divergence of
$f$ for $\beta < 0$ will show up for $\alpha < \alphavctyp$ 
\footnote{For binary patterns the probability for two identical
    patterns is $2^{-N}$ and $f(\beta)$ should be divergent for
    $\beta<0$ for all $\alpha$. This is, however, not found in the
    explicit calculation since by keeping only the first two moments
    of the pattern distribution in performing the ensemble average
    one effectively replaces the original distribution by a Gaussian
    one.}. 
For $\alpha > \alphavctyp$, however, there are {\em
typically} some empty cells and $f (\beta)$ should be divergent for all
$\beta < 0$.

Within the replica symmetric approximation one finds this divergence of
negative moments for both the spherical and the Ising perceptron at 
$\beta =(\alpha - 1)/\alpha$ if $\alpha < 1$ and $\beta = 0^{-}$ if 
$\alpha\geq 1$ \cite{engel:weigt:96}. This suggests $\alphavctyp = 1$ for both cases.
For the spherical perceptron this coincides with the known result. Moreover
the point $\beta = 0, \alpha = 1$ belongs to the region of local stability
of the replica symmetric saddle point. For the Ising case the result must be
wrong since $\alphavctyp$ cannot be larger than $\alpha_c \approx 0.83$. 
Since also in this case the replica symmetric saddle point is locally stable
at $\beta = 0, \alpha = 1$ it is very likely that there is a discontinuous
transition to replica symmetry breaking as typical for this system
\cite{krauth:mezard:89}. It remains to be seen whether a solution in one step replica
symmetry breaking can provide a more realistic value of $\alphavctyp$.

In principle it is possible using the same techniques to obtain expressions for the 
typical growth function of simple multi-layer nets. However, the technical problems will 
increase and replica symmetry breaking is again likely to show up. We just note that a related
analysis, namely the characterization of the distribution of internal representations
within the typical Gardner volume has recently been performed 
\cite{monasson:zecchina:95,monasson:zecchina:96,cocco:monasson:zecchina:96} for the
committee machine. From these investigations a new result for the storage capacity in the limit
of a large number of hidden units was obtained.

\section{Typical growth functions}
\label{sec:typical}

The typical growth function $\Deltatyp(p)$ of
a classifier system that is parameterized by $N$ binary variables can be
measured numerically by an algorithm that mixes Monte-Carlo methods
and exact enumeration \cite{stambke:92}.

The enumeration is required to determine 
$\Delta(\boldsymbol{\xi}^1,\ldots,\boldsymbol{\xi}^p)$, the number
of different output vectors that are realizable for a given pattern set.
To get this number, one has to calculate the output vectors of all
$2^N$ classifiers! This exponential complexity limits the numerical
calculations to small values of $N$.

To get $\Deltatyp(p)$, we draw $p$ random unbiased patterns $\boldsymbol{\xi}^\mu\in\{\pm1\}^N$
and calculate $\Delta(\boldsymbol{\xi}^1,\ldots,\boldsymbol{\xi}^p)$. This
is repeated again and again, and the values of $\Delta(\boldsymbol{\xi}^1,\ldots,\boldsymbol{\xi}^p)$
are averaged to yield $\Deltatyp(p)$. The scale of $\Delta$ for large
$N$ is $O(2^N)$ so we average the logarithm:
\begin{equation}
  \label{eq:Delta_average}
  \log \Deltatyp(p) = \left<\log(\Delta(\boldsymbol{\xi}^1,\ldots,\boldsymbol{\xi}^p))\right>_{\{
  \boldsymbol{\xi}^\mu\}}
\end{equation}

\begin{figure}[htb]
  \includegraphics[width=\columnwidth]{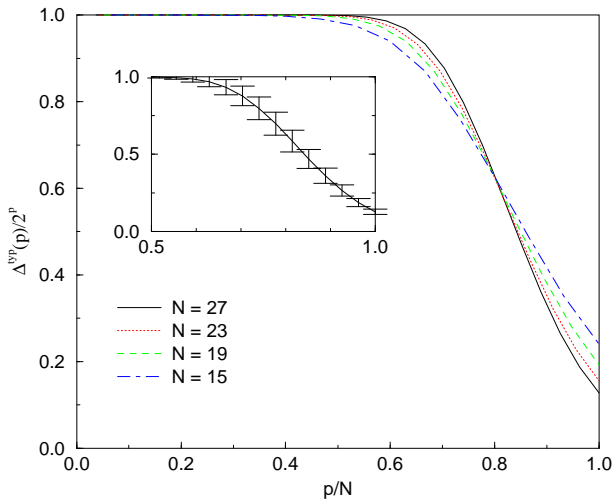}
        \caption[Fig1]{\label{fig:percdich}
           Typical growth function of the Ising perceptron with binary patterns averaged over 1000 samples. 
           The function
           is of course only defined at discrete values of $p$, but the continuous lines ease the
           readability. The inset displays the values for $N=27$ together with the error bars.
        }
\end{figure}

Figure \ref{fig:percdich} shows $\Deltatyp(p)$ for the Ising perceptron with binary patterns. The
curves display the expected behavior: $\Deltatyp(p) = 2^p$ for small $p$ and $\Deltatyp(p)\ll 2^p$
for larger values of $p$. The transition between these to regimes seems to become sharper with increasing
$N$, but it is not clear whether we get a true step-function in the limit $N\to\infty$. The corresponding
curves for the committee- and the parity-tree look similar.

\begin{figure}[htb]
  \includegraphics[width=\columnwidth]{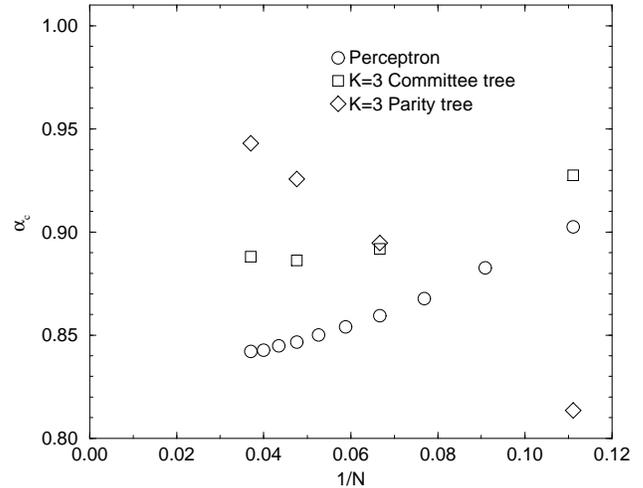}
        \caption[Fig1]{\label{fig:alpha_c}
           Critical storage capacity $\alpha_c$ deduced from $\Deltatyp(p)$ for the Ising
           perceptron, the $K=3$ Ising committee tree and the $K=3$ Ising parity tree.
        }
\end{figure}

As a test we derive the critical storage capacity $\alpha_c$ from fig.~\ref{fig:percdich}
by reading off the point where $\Deltatyp(p)=2^{p-1}$.
Figure \ref{fig:alpha_c} shows $\alpha_c$ vs.\ $1/N$ for the Ising perceptron and the committee-
and parity-tree with $K=3$ each. The extrapolations to $N=\infty$ are in good agreement with the anaytical
results $\alpha_c=0.83$ for the Ising perceptron \cite{krauth:mezard:89}, 
$\alpha_c=0.92$ for the Ising committee tree with $K=3$ \cite{barkai:hansel:sompolinsky:92} and
$\alpha_c=1$ for the Ising parity tree with $K\geq 2$ \cite{barkai:kanter:90}.

For the spherical perceptron $\Delta(\{\boldsymbol{\xi}^\mu\})$ is known to be the same
for all pattern sets in general position. The inset of fig.~\ref{fig:percdich} displays
that in the case of the Ising perceptron the average over the patterns introduces a statistical
error that does \emph{not} tend to zero with increasing number of samples. This implies that for the
Ising perceptron the number of realizable output sequences is not the same for all pattern
sets in general position.

\begin{figure}[htb]
  \includegraphics[width=\columnwidth]{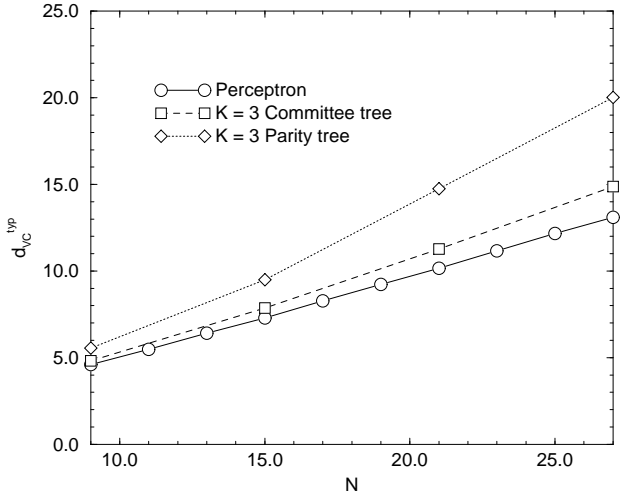}
        \caption[Fig1]{\label{fig:vc_typ}
           Numerical values of $\dvct(N)$ for the Ising
           perceptron, the $K=3$ Ising committee tree and the $K=3$ Ising parity tree.
           The straight lines between the points are guides to the eye.
        }
\end{figure}

The typical VC-dimension $\dvct$ can principally be obtained from $\Deltatyp(p)$ as the
number of patterns for which $\Deltatyp(p)$ starts to deviate from $2^p$. Due to the
statistical errors in $\Deltatyp(p)$, a separate evaluation of $\dvct$ is more appropriate.
For this, we calculate $\Delta(\boldsymbol{\xi}^1,\ldots,\boldsymbol{\xi}^p)$ for a random
set of patterns. If equal to $2^p$, the set is enlarged by another random pattern and
$\Delta(\{\boldsymbol{\xi}^\mu\})$ is calculated again. This step is repeated until
the set is no longer shattered. The number of patterns in the set (minus 1) gives
a value for $\dvct$. These values are averaged over many random samples. The results
are shown in fig.~\ref{fig:vc_typ}.
The dependence of $\dvct$ on $N$ is roughly given by
\begin{equation}
  \begin{split}
  \label{eq:dvctyp}
  \text{ Ising perceptron: }& \qquad \dvct(N) \propto \frac12\, N \\
  \text{$K=3$ committee-tree: }& \qquad \dvct(N) \propto 0.6\, N \\
  \text{$K=3$ parity-tree: }& \qquad \dvct(N) \propto 0.88\, N.
  \end{split}
\end{equation}

\section{Bounds for $\dvc$}
\label{sec:exact}

The exact value of $\dvc$ for the Ising perceptron with binary or zero threshold 
and binary patterns is not known, not even in the limit $N\to\infty$. Only bounds can be provided.

An arbitrary set of classifiers which are parameterized by $N$ bits -- like the  
Ising perceptron -- cannot produce more than $2^N$
distinct output vectors on any set of input patterns. So we have
\begin{equation}
  \label{eq:upper_bound}
  \dvc(N) \leq N
\end{equation}
as a general upper bound for ``Ising-like'' classifiers.

Finding good lower bounds is a bit more tedious. It can be achieved 
by explicit construction of shattered sets. In those cases, 
however, where $\dvct \ll \dvc$, shattered sets with cardinality $>\dvct$ are rare and
consequently hard to find by random search.

\subsection{Ising perceptron}

For the Ising perceptron it is shown in appendix \ref{sec:symmetry} that $\dvc$ is the same
for $N$ odd, zero threshold and $N-1$, binary threshold. Therefore we can safely restrict ourselves
to the case $N$ odd and no threshold.

In ref.~\cite{mertens:96a} a special pattern set is given that yields
\begin{equation}
  \label{eq:low_bound_3}
  \dvc(N) \geq \frac12 (N+3).
\end{equation}
Shattered sets with cardinality $\frac12 (N+3)$ are not too rare; they do show up in the statistical
algorithm of section \ref{sec:typical}.

To get an improved lower bound for general values of $N$, we consider a restricted
variant of the Ising perceptron, the {\em balanced} Ising perceptron where
the couplings have minimum ``magnetization'':
\begin{equation}
   \label{eq:balanced}
   \sum_i J_i=\pm1
\end{equation}
The balanced Ising perceptrons are a subset of the usual Ising perceptrons,
hence any pattern set that is shattered by the former is as well
shattered by the latter.
 
Now let $\{\boldsymbol{\xi}^1,\ldots,\boldsymbol{\xi}^p\}$ be a shattered set
for the balanced Ising perceptron with $N$ nodes and  let $\{\sigma_1,\ldots,\sigma_p\}$ be an
output vector that is realized by the balanced weight-vector $\mathbf{J}$.
Going from $N$ to $N+2$, we define $p+1$ patterns\ldots
\begin{equation}
\label{eq:pattern_trafo}
\begin{split}
  \boldsymbol{\tilde\xi}^\nu & = (-, \boldsymbol{\xi}^\nu, -) \quad 1 \leq \nu \leq p\\
  \boldsymbol{\tilde\xi}^{p+1} & = (-, \underbrace{-,\cdots,-}_{N \mathrm{times}},+)
\end{split}
\end{equation}
\ldots and new couplings
\begin{equation}
\label{eq:coupl_trafo}
\begin{split}
  \mathbf{J}^{\pm} &= (+, \mathbf{J}, -)\\
  \mathbf{J}^{\mp} &= (-, \mathbf{J}, +).
\end{split}
\end{equation}
These couplings preserve the output values of the ``old'' patterns
\begin{equation}
\sign(\mathbf{J}^{\pm}\boldsymbol{\tilde\xi}^\nu) = \sign(\mathbf{J}^{\mp}\boldsymbol{\tilde\xi}^\nu) = 
       \sigma_\nu \quad 1\leq \nu \leq p,
\end{equation}
while the balance property ensures that both classifications of the new pattern can be realized:
\begin{equation}
\begin{split}
\sign(\mathbf{J}^{\pm}\boldsymbol{\tilde\xi}^{p+1}) &= \sign(-2 - \sum_{i=1}^NJ_i) = -1\\
\sign(\mathbf{J}^{\mp}\boldsymbol{\tilde\xi}^{p+1}) &= \sign(2 - \sum_{i=1}^NJ_i) = +1.
\end{split}
\end{equation}
Note that $\mathbf{J}^{\pm}$ and $\mathbf{J}^{\mp}$ both are balanced.
This allows us to apply eqs.~(\ref{eq:pattern_trafo},\ref{eq:coupl_trafo}) recursively to obtain
the lower bound
\begin{equation}
  \label{low_bound_c}
  \dvc(N) \geq \frac12 (N + 2c - N_0) \quad N\geq N_0
\end{equation}
for the \emph{general} Ising percpetron,
where $c$ is given by the cardinality of a shattered set for the \emph{balanced} Ising perceptron
with $N_0$ nodes.

Now we are left with the problem of finding large shattered sets for the balanced Ising perceptron.
A partial enumeration (see below) yields shattered sets with cardinality $c=7,11,13$ for 
$N=9,15,17$. This gives
\begin{equation}
\label{eq:low_bounds_balanced}
\begin{split}
\dvc(N) &\geq \frac12(N+5) \quad N\geq 9 \\
\dvc(N) &\geq \frac12(N+7) \quad N\geq 15 \\
\dvc(N) &\geq \frac12(N+9) \quad N\geq 17
\end{split}
\end{equation}
The corresponding shattered sets are listed in appendix \ref{sec:gallery}. This sequence
of increasing lower bounds indicates that probably $\lim_{N\to\infty}\frac{\dvc(N)}N > \frac12$.

There is a method that surely finds the largest possible shattered set, i.e.\ the exact value of
$\dvc$: \emph{exhaustive enumeration} of all shattered sets. The overwhelming complexity of
$O(2^{N^2})$ limits this approach to small values of $N$, however.
Nevertheless, the results obtained for $N\leq 9$ are already quite remarkable  \cite{mertens:96a}:
\begin{equation}
  \label{eq:exact_dvc}
  \begin{split}
    \dvc(3) &= 3 \\
    \dvc(5) &= 4 \\
    \dvc(7) &= 7 \\
    \dvc(9) &= 7 
  \end{split}
\end{equation}
Again the corresponding shattered sets are listed in appendix \ref{sec:gallery}.
They share a common feature: Using transformations that do not change $\Delta(p)$
(see appendix \ref{sec:gallery}), they can be transformed into quasi orthogonal 
pattern sets, i.e.\ into sets, where the patterns have minimum pairwise 
overlap\footnote{Exact orthogonality 
cannot be achieved for $N$ odd.}
\begin{equation}
  \label{eq:orthogonality}
  \boldsymbol{\xi}^{(\mu)}\cdot\boldsymbol{\xi}^{(\nu)} = \left\{
  \begin{array}{rr}
     \pm 1 & \mu\neq\nu\\
                 N & \mu=\nu
  \end{array}
  \right..
\end{equation}
\begin{figure}[htb]
  \includegraphics[width=\columnwidth]{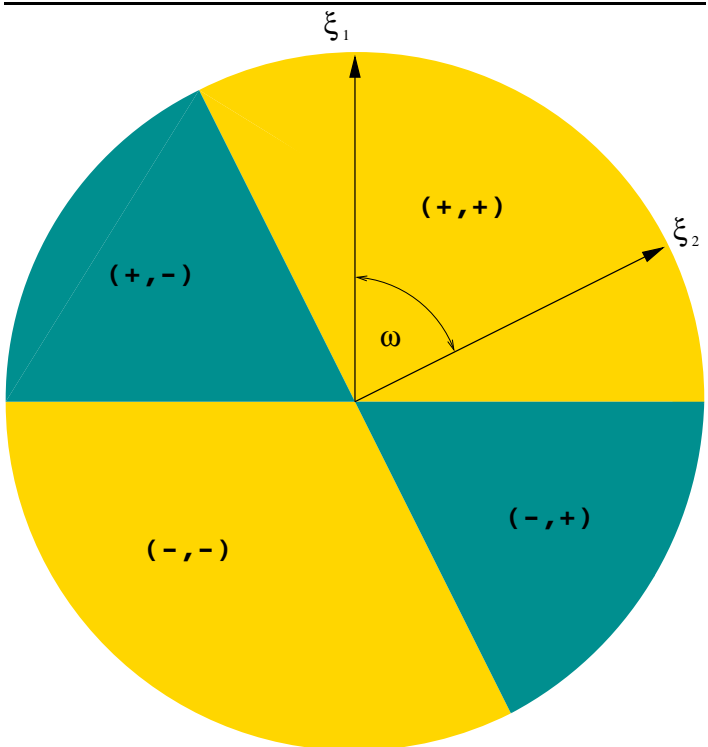}
        \caption[Fig1]{\label{fig:ortho}Spherical perceptron with $N=2$. The 4 cells in weigth-space
induced by patterns $\boldsymbol{\xi}_1$ and $\boldsymbol{\xi}_2$ are equisized for $\omega=\pi/2$, i.e.\ for
orthogonal patterns.}
\end{figure}

This observation appears reasonable: Consider a shattered set of patterns. The corresponding cells
in weight-space have non-zero volume $V(\{\sigma_\mu\})$, i.e. each cell contains at least one weight vector
$\mathbf{J}$. If we enlarge the shattered set by an additional pattern, each cell must divide in two
cells of non-zero volume. This process can be repeated until the first non divisible cell appears.
If we assume that the divisibility of a cell decreases with its volume, we must look for cell structures
where the volume of the smallest cell is maximized. This is the case for \emph{equisized} cells, i.e.\ for
orthogonal patterns (fig.~\ref{fig:ortho}).

\begin{figure}[htb]
  \includegraphics[width=\columnwidth]{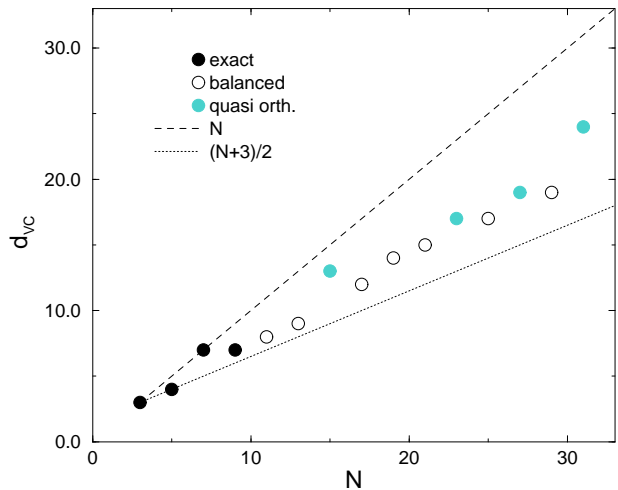}
        \caption[Fig1]{\label{fig:vc}VC dimension of the Ising perceptron with binary patterns vs.\ $N$.
The circles labeled 'quasi orthogonal' and 'balanced' are lower bounds for the true $\dvc$.
}
\end{figure}

Quasi orthogonal pattern sets can easily be built from the rows of Hadamard matrices 
(see appendix \ref{sec:hadamard}). These are $4n\times 4n$ orthogonal matrices with
$\pm1$ entries. To get quasi orthogonal patterns of odd length $N$, we either cut out
one column ($N=4n-1$) or add an arbitrary column ($N=4n+1$). It is clear that there are
many quasi orthogonal pattern sets with $p$ elements that can be constructed from a 
given Hadamard matrix. By \emph{partial enumeration}, i.e.\ by evaluation
of some of them, we were able to find shattered sets that exceed the lower bound given by
eq.~(\ref{eq:low_bounds_balanced}) for certain values of $N$:
\begin{equation}
\label{eq:low_bounds_hadamard}
\begin{split}
\dvc(N=15) &\geq 13\\
\dvc(N=23) &\geq 17\\
\dvc(N=27) &\geq 19\\
\dvc(N=31) &\geq 24
\end{split}
\end{equation}
The corresponding pattern sets are listed in appendix \ref{sec:gallery}. Systems with $N > 31$ were not
investigated. Note that the lower bound for $\dvc(N=31)$ is larger than the value reported in 
ref.~\cite{mertens:96a}.

Fig.~\ref{fig:vc} summarizes our results for the $N\leq 31$. Both the exact values and the lower
bounds provided by eq.~(\ref{eq:low_bounds_balanced}) and eq.~(\ref{eq:low_bounds_balanced}) clearly
exceed the maximum value $\dvc=\frac12(N+3)$ found by the statistical method in section \ref{sec:typical}.
The somewhat irregular behaviour of the lower bounds does not rule out a more regular sequel of the
true $\dvc(N)$, including well defined asymptotics. However, if the limit $\lim_{N\to\infty}\dvc/N$
exists, it will probably be larger than $0.5$.

\subsection{Committee tree}

To get a lower bound for $\dct$, the VC-dimension of the Ising committee tree with binary patterns,
we explicitely construct a shattered set based on shattered sets for the Ising perceptron.

Let $\{\boldsymbol{\tau}^\nu\}$ be a shattered set of an Ising perceptron with $K$ nodes and
$\{\boldsymbol{\xi}^\mu\}$ be a shattered set of the $\frac NK$-node subperceptron.
Then we build patterns $\boldsymbol{\Xi}^{\mu,\nu}$ for the committee tree by concatenation
\begin{equation}
  \label{eq:cartesian_set}
  \boldsymbol{\Xi}^{\mu,\nu} = 
      (\tau_1^\nu\boldsymbol{\xi}^\mu, \tau_2^\nu\boldsymbol{\xi}^\mu, \ldots, \tau_k^\nu\boldsymbol{\xi}^\mu).
\end{equation}
We proof, that the set $\{\boldsymbol{\Xi}^{\mu,\nu}, 1\leq\mu\leq\dvc(N/K), 1\leq\nu\leq\dvc(K)\}$ is shattered.

Let $\{\sigma_{\mu,\nu}\}$ be a given output sequence of length $\dvc(K)\dvc(N/K)$. Since
the $\{\boldsymbol{\tau}^\nu\}$ are shattered, we can always find $\{W_k^\mu=\pm1\}_{k=1}^K$ such that
\begin{equation}
  \label{eq:output_layer}
  \sigma_{\mu,\nu} = \sign\left(\sum_{k=1}^K W_k^\mu\tau_k^\nu\right)
\end{equation}
for all $\nu$ and $\mu$.
Now we choose the couplings in the $k$-th subperceptron such that
\begin{equation}
  \label{eq:hidden_layer}
  W_k^\mu = \sign\left(\sum_{i=1}^{\frac NK}J_i^{(k)} \xi_i^\mu\right) \quad k=1,\ldots,K.
\end{equation}
This is always possible because $\boldsymbol{\xi}^\mu$ is taken from a shattered set.
Combining eqs.\ (\ref{eq:output_layer}) and (\ref{eq:hidden_layer}) we get
\begin{equation}
\sigma_{\mu,\nu} = \sign\left(\sum_{k=1}^K\sign\sum_{i=1}^{\frac NK}J_i^{(k)}\tau_k^\nu\xi_i^\mu
               \right) \quad \nu=1,\ldots,\dvc(K).
\end{equation}
i.e.\ the patterns (\ref{eq:cartesian_set}) form a shattered set and we find
\begin{equation}
  \label{eq:low_bound_ct}
  \dct(N) \geq \dvc(K) \dvc(N/K)
\end{equation}
Note that this lower bound matches the upper bound $N$ whenever $\dvc(K)$ and $\dvc(N/K)$ meet their upper
bounds $K$ resp.\ $N/K$. Examples: $K=3$ or $K=7$ and $N=21$, $K=7$ and $N=49$.

This lower bound is much larger than the values for $\dvct$ found section~\ref{sec:typical}. If we assume
that $\alphavc = \lim_{N\to\infty}\dvc(N)/N$ is well defined for the Ising percpetron,
eq.~(\ref{eq:low_bound_ct}) reads
\begin{equation}
  \label{eq:low_bound_alpha_ct}
  \dct(N) \geq N\alphavc^2 \qquad N>>K>>1
\end{equation}

\subsection{Parity tree}

We follow the same strategy and construct a shattered set from the patterns of a shattered
set $\{\boldsymbol{\xi}^\nu\}$ of the subperceptrons. The first pattern is simply built from $K$
consecutive patterns $\boldsymbol{\xi}^1$:
\begin{equation}
  \boldsymbol{\Xi}^0 = (\boldsymbol{\xi}^1,\ldots,\boldsymbol{\xi}^1)
\end{equation}
All other patterns differ from $\boldsymbol{\Xi}^0$ in only one subpattern:
\begin{equation}
  \begin{split}
  \boldsymbol{\Xi}^{\mu,k} = (\boldsymbol{\xi}^1,\ldots,\boldsymbol{\xi}^1,&\boldsymbol{\xi}^\mu,
                              \boldsymbol{\xi}^1,\ldots,\boldsymbol{\xi}^1)\\
  & \uparrow \mbox{$k$-th position}
  \end{split}
\end{equation}
where $k = 1,\ldots,K$ and $\mu = 2,\ldots,\dvc(N/K)$. This set of $k(\dvc(N/K)-1) + 1$ 
patterns is shattered. 

Proof: Let $\{\sigma_0,\sigma_{k,\mu}\}$ be a given output sequence for our patterns. We choose
the weights $\mathbf{J}^{(k)}$ in the subperceptrons such that
\begin{equation}
\label{eq:weights_parity}
\begin{split}
\sign(\mathbf{J}^{(1)}\cdot\boldsymbol{\xi}^1) &= \sigma_0 \\
\sign(\mathbf{J}^{(k>1)}\cdot\boldsymbol{\xi}^1) &= 1 \\
\sign(\mathbf{J}^{(1)}\cdot\boldsymbol{\xi}^\mu) &= \sigma_{1,\mu} \\
\sign(\mathbf{J}^{(k>1)}\cdot\boldsymbol{\xi}^\mu) &= \sigma_0\sigma_{k,\mu}
\end{split}
\end{equation}
for $\mu = 2\ldots\dvc(N/K)$. This is always possible because $\{\boldsymbol{\xi}^\nu\}$
is shattered. With this assignment of weights, the parity tree maps
$\{\boldsymbol{\Xi}^0,\boldsymbol{\Xi}^{\mu,n}\}$ to the prescribed output sequence.

Our shattered set provides us with a lower bound for the VC-dimension of the parity tree
\begin{equation}
  \label{eq:low_bound_pt}
  \dpt(N) \geq K(\dvc(N/K)-1) + 1.
\end{equation}
For $K=1$ the parity tree is equivalent to the simple perceptron and eq.~(\ref{eq:low_bound_pt})
reduces to $\dpt(N) = \dvc(N)$. If one inserts the lower bounds for $\dvc$ into the r.h.s.\ 
of eq.~\ref{eq:low_bound_pt}, the resulting values are generally larger than $\dvct$, but
the differences are much smaller than for the committee-tree, and for some values of $N$ 
$\dvct$ even exceeds the r.h.s.\ of eq.~\ref{eq:low_bound_pt}. We do not know whether
 eq.~\ref{eq:low_bound_pt} is only a bad lower bound or whether the maximum shattered sets
for the parity-tree are not as atypical as for the Ising perceptron and the committee-tree.

\section{Conclusions}
\label{sec:coda}

The VC-dimension is one of the central quantities to characterize
the information processing abilities of feed-forward neural networks.
The determination of the VC-dimension of a given network architecture 
is, however, in general a non-trivial task.

In the present paper we have shown that even for the simplest feed-forward
neural networks this task requires rather sophisticated techniques if both
the couplings of the network and the inputs are restricted to binary values
$\pm1$. This is mainly due to the fact that the VC-dimension defined by a 
suprenum over all pattern sets of given size is determined by \emph{atypical}
pattern sets. Consequently Monte-Carlo methods as well as analytical estimates 
involving pattern averages do not yield reliable results and one has to resort
to exact enumeration techniques. These methods are naturally restricted to small
dimensions of the input space but the results obtained can be used to get lower
bounds for the VC-dimension of larger systems. In some cases, even higher bounds
can be derived from number theoretic arguments.

Complementary one could argue that \emph{typical} situations are of more
interest than the worst case. Accordingly a typical VC-dimension $\dvct$ has been
defined in section~\ref{sec:basics}. One always has $\dvct\leq\dvc$ since
an average can never be larger than the suprenum.

For the Ising perceptron ($J_i=\pm1$) we found $\dvc=N$ as long as the patterns are
allowed to take on the value 0, no matter whether we use real-valued
or $\{0,1\}$-patterns. 
If, however, also the patterns are Ising-like, i.e.\ $\boldsymbol{\xi}\in\{\pm1\}^N$ our
numerical results suggest
\begin{equation}
  \label{eq:conclusion}
  \frac12(N+3) < \dvc(N) < N
\end{equation}
for general $N$. 
For large $N$, the VC-dimension is presumably \emph{substantially} larger than the 
typical VC-dimension $\dvct(N)\propto N/2$.

Similar results are found for two simple examples of multi-layer networks,
the committee- and the parity-tree with Ising couplings. Here the results
are
\begin{equation}
  \dvc(K) \dvc(N/K) \leq \dct(N) \leq N
\end{equation}
for the committee- and
\begin{equation}
  \label{eq:xxx}
  K(\dvc(N/K)-1)+1 \leq \dpt(N) \leq N
\end{equation}
for the parity-tree with $K$ hidden nodes.
For the committee-tree we find again that $\dvct < \dvc$. For the parity-tree, our 
data does not allow to draw the same conclusion, but this may be due to the low
quality of the lower bound in eq.~(\ref{eq:xxx}).

We finally note that the growth function $\Delta(p)$ related to the VC-dimension is used to
derive the famous Vapnik-Chervonenkis bound for the asymptotic difference between learning
and generalization error. This bounds results from the analysis of the worst case. It would be
interesting to investigate whether a similar bound for the \emph{typical} generalization
behaviour could be obtained from $\Deltatyp(p)$ which in general is much easier to
determine.

\appendix
\section{Symmetries}
\label{sec:symmetry}

Let $\{\boldsymbol{\xi}^1,\ldots,\boldsymbol{\xi}^p\}$ be a 
set of binary $\pm1$ patterns and $\Delta(\boldsymbol{\xi}^1,\ldots,\boldsymbol{\xi}^p)$
the number of different output sequences  $(\sigma_1,\ldots,\sigma_p)$ that
can be realized by the Ising-perceptron for this particular set of patterns.

$\Delta(\boldsymbol{\xi}^1,\ldots,\boldsymbol{\xi}^p)$ is invariant under the
following transformations on $\{\boldsymbol{\xi}^1,\ldots,\boldsymbol{\xi}^p\}$:
\begin{itemize}
\item complement a whole pattern: $\boldsymbol{\xi}^\mu\mapsto-\boldsymbol{\xi}^\mu$
\item interchange two patterns: $\boldsymbol{\xi}^\mu \leftrightarrow \boldsymbol{\xi}^\nu$
\item complement one entry in all patterns ($\mu=1\ldots p$): $\xi_i^\mu \mapsto -\xi_i^\mu$
\item interchange two entries in all patterns ($\mu=1\ldots p$): $\xi_i^\mu \leftrightarrow \xi_j^\mu$
\end{itemize}
Applying these transformations, we can always achieve that all patterns have $\xi_N^\mu = -1$.

Now we assume that $N$, the number of couplings, is odd and that there
is no threshold. Let $\mathbf{J}$ be a weight vector that realizes
an output sequence $(\sigma_1,\ldots,\sigma_p)$ for a patterns with
$\xi_N^\mu=-1$:
\begin{equation}
\label{odd_even}
\sigma_\nu = \sign(\sum_{i=1}^{N-1}J_i\xi_i^\nu - J_N)\quad 1 \leq \nu \leq p.
\end{equation}
We use the left $N-1$ bits of $\boldsymbol{\xi}^\nu$ as a pattern set for the Ising-perceptron
with $N-1$ input units and a binary threshold $\Theta$.
Identifying $\Theta$ with $J_N$ in eq.~(\ref{odd_even}) it becomes obvious that 
$\Delta(\boldsymbol{\xi}^1,\ldots,\boldsymbol{\xi}^p)$ is the same in both cases.
Hence we may restrict ourselves to the case
$N$ odd and no threshold to discuss the VC-dimension of the Ising-perceptron.

Now we consider a 2 layer feedforward network with $K$ perceptrons (spherical or Ising)
operating between input and hidden layer (weight vectors $\mathbf{J}^{(k)}$
and an Ising perceptron as decoder function with weight vector $\mathbf{J}^{(0)}$.
Suppose that a given output sequence is realized by a weight vector with
some entries $J_k^0 = -1$ in the decoder perceptron.
The output sequence is left unchanged if we set $J_k^0 = +1$ and at the same
time complement all weights in the $k$-th subperceptron $\mathbf{J}^{(k)}\mapsto -\mathbf{J}^{(k)}$.
This transformation allows us to realize any realizable output sequence with all $J_k^0=+1$.
Hence the VC-dimension of the committee machine equals the VC-dimension of the two layer
perceptron with Ising weights in the output layer.

\section{Hadamard matrices}
\label{sec:hadamard}

A Hadamard matrix is an
$m\times m$-matrix $H$ with $\pm1$-entries such that 
\begin{equation}
  \label{hadamard}
  HH^T = mI
\end{equation}
where $I$ is the $m\times m$ identity matrix. 

If $H$ is an $m\times m$ Hadamard matrix, then $m=1$, $m=2$
or $m\equiv0\bmod4$. The reversal is a famous open question: Is there a Hadamard
matrix of order $m=4n$ for every positive $n$? The first open case is
$m=428$.

If $H$ and $H'$ are Hadamard matrices of order $m$ resp.\ $m'$, their Kronecker 
product $H\otimes H'$ is a Hadamard matrix of order $mm'$. Starting with the
$2\times2$ Hadamard matrix
\begin{equation}
 H_2 = \left(
\begin{array}{cc}
  -1 & -1 \\
        -1 & +1
\end{array}
\right).
\end{equation}
this gives Hadamard matrices of order $4,8,16,\ldots,2^n$, the so called \emph{Sylvester type}
matrices. Example:
\begin{equation}
  \label{eq:H8}
  H_{2^3} = \left(
    \begin{smallmatrix}
       -1 & -1 & -1 & -1 & -1 & -1 & -1 & -1 \\
       -1 & +1 & -1 & +1 & -1 & +1 & -1 & +1 \\
       -1 & -1 & +1 & +1 & -1 & -1 & +1 & +1 \\
       -1 & +1 & +1 & -1 & -1 & +1 & +1 & -1 \\
       -1 & -1 & -1 & -1 & +1 & +1 & +1 & +1 \\
       -1 & +1 & -1 & +1 & +1 & -1 & +1 & -1 \\
       -1 & -1 & +1 & +1 & +1 & +1 & -1 & -1 \\
       -1 & +1 & +1 & -1 & +1 & -1 & -1 & +1 
    \end{smallmatrix}
  \right)
\end{equation}

Let $q$ be an odd prime power. Then Hadamard matrices of \emph{Paley type} 
can be constructed for
\begin{equation}
  m = \left\{
  \begin{array}{lr}
    q+1 & \text{ for } q\equiv3\bmod4 \\
   2(q+1) & \text{ for } q\equiv1\bmod4 
  \end{array}
  \right.
\end{equation}

Paley's construction \cite{beth:etal:85} relies on the properties of finite Galois fields $\mathrm{GF}(q)$
\cite{lidl:niederreiter:94}, where
$q$ is an odd prime power, especially on the \emph{quadratic character} $\chi$ of $\mathrm{GF}(q)$,
defined by
\begin{equation}
  \label{eq:def_chi}
  \chi(x) = \left\{
  \begin{array}{rl}
    0 & \text{ if $x=0$}\\
    +1 & \text{ if $x\neq0$ is a square}\\
    -1 & \text{ otherwise. }
  \end{array}
  \right.
\end{equation}
Then for any $a\neq 0$
\begin{equation}
  \label{eq:chi_correlation}
  \sum_{x\in\mathrm{GF}(q)} \chi(x)\chi(x-a) = -1.
\end{equation}

To construct a Paley-type matrix for $q\equiv3\bmod4$, we start with
the $q\times q$ matrix $M = (m_{ij})$ whose rows and columns are indexed
by the elements of $\mathrm{GF}(q)$:
\begin{equation}
  m_{ij} = \left\{
  \begin{array}{lll}
    -1 & \text{ if } & i = j\\
    \chi(i-j) & \text{ if } &i\neq j.
  \end{array}
  \right.
\end{equation}
Hence by eq.~(\ref{eq:chi_correlation})
\begin{equation}
\sum_{j\in\mathrm{GF}(q)} m_{hj}m_{ij} = \left\{
\begin{array}{rl}
  q & h = i\\
  -1 & h \neq i.
\end{array}
\right.
\end{equation}
Now adjoin one row and one column with all entries $+1$ to get a Hadamard matrix
of order $q+1$. This gives Hadamard matrices of order $4,8,12,20,24,28,\ldots$.

Example: $q=11$. The Galois field $\mathrm{GF}(11)$ is equivalent to the integers $\{0,\ldots,10\}$
together with their addition and multiplication modulo 11. The squares are 1,4,9,5,3 and we get
\begin{equation}
  \label{eq:H11p1}
  \setcounter{MaxMatrixCols}{13}
  H_{11+1} = \left(
    \begin{smallmatrix}
      +1 & +1 & +1 & +1 & +1 & +1 & +1 & +1 & +1 & +1 & +1 & +1\\
      +1 & -1 & +1 & -1 & +1 & +1 & +1 & -1 & -1 & -1 & +1 & -1\\
      +1 & -1 & -1 & +1 & -1 & +1 & +1 & +1 & -1 & -1 & -1 & +1\\
      +1 & +1 & -1 & -1 & +1 & -1 & +1 & +1 & +1 & -1 & -1 & -1\\
      +1 & -1 & +1 & -1 & -1 & +1 & -1 & +1 & +1 & +1 & -1 & -1\\
      +1 & -1 & -1 & +1 & -1 & -1 & +1 & -1 & +1 & +1 & +1 & -1\\
      +1 & -1 & -1 & -1 & +1 & -1 & -1 & +1 & -1 & +1 & +1 & +1\\
      +1 & +1 & -1 & -1 & -1 & +1 & -1 & -1 & +1 & -1 & +1 & +1\\
      +1 & +1 & +1 & -1 & -1 & -1 & +1 & -1 & -1 & +1 & -1 & +1\\
      +1 & +1 & +1 & +1 & -1 & -1 & -1 & +1 & -1 & -1 & +1 & -1\\
      +1 & -1 & +1 & +1 & +1 & -1 & -1 & -1 & +1 & -1 & -1 & +1\\
      +1 & +1 & -1 & +1 & +1 & +1 & -1 & -1 & -1 & +1 & -1 & -1
    \end{smallmatrix}
  \right)
  \setcounter{MaxMatrixCols}{10}
\end{equation}

For $q\equiv1\bmod4$, the construction starts with the $(q+1)\times(q+1)$ Matrix $M=(m_{ij})$,
indexed by $\mathrm{GF}(q)\cup\{\infty\}$ as follows
\begin{equation}
\begin{split}
m_{\infty j} = m_{j\infty} &= 1 \text{ for all $j\in\mathrm{GF}(q)$ } \\
m_{\infty\infty} &= 0 \\
m_{ij} &= \chi(j-i) \text{ for $i,j \in \mathrm{GF}(q)$. }
\end{split}
\end{equation}
$M$ is symmetric and orthogonal. To get from $M$ to a Hadamard matrix of order
$2(q+1)$, we define the auxiliary matrices $A$ and $B$ by
\begin{equation}
  A = \left(
  \begin{array}{rr}
    1 & 1 \\
    1 & -1 
  \end{array}
  \right)
  \quad
  B = \left(
  \begin{array}{rr}
    1 & -1 \\
   -1 & -1 
  \end{array}
  \right)
\end{equation}
and replace every 0 in $M$ by $B$, every $+1$ by $A$ and every $-1$ by $-A$.
This gives Hadamard matrices of order $12,20,28,36,52,\ldots$.
Example: $q=5$. $\mathrm{GF}(5)$ is equivalent to the integers $\{0,\ldots,4\}$
and their addition and multiplication modulo 5. The squares are 1 and 4 and we get
\begin{equation}
  \label{eq:H2m5p1}
  H_{2(5+1)} = \left(
    \begin{array}{rrrrrr}
       B & A & A & A & A & A \\
       A & B & A & -A & -A & A \\
       A & A & B & A & -A & -A \\
       A & -A & A & B & A & -A \\
       A & -A & -A & A & B & A \\
       A & A & -A & -A & A & B 
    \end{array}
  \right)
\end{equation}

The first value of
$m = 4n$ where neither the Sylvester- nor the Paley-construction applies is $m = 92$.

\section{Gallery of shattered sets}
\label{sec:gallery}
For $N \leq 9$ the exact values of $\dvc$ have been obtained by exhaustive
enumerations. Shattered sets of maximum cardinality are:

\begin{tabular}{cccc}
    $N=3$ &       $N=5$ &         $N=7$ &           $N=9$ \\
\pat{+++} & \pat{+++++} & \pat{-------} & \pat{---------} \\
\pat{--+} & \pat{-+--+} & \pat{-+-+-+-} & \pat{+-+-+-+-+} \\
\pat{-+-} & \pat{-+-+-} & \pat{--++--+} & \pat{---++--++} \\
          & \pat{-++--} & \pat{-++--++} & \pat{+-++--++-} \\
          &             & \pat{----+++} & \pat{-----++++} \\
          &             & \pat{-+-++-+} & \pat{+-+-++-+-} \\
          &             & \pat{--++++-} & \pat{---++++--}
\end{tabular}

The sets for $N=3$ and $N=5$ are obtained from the rows of the Sylvester type
Hadamard matrix $H_{2^2}$. For $N=3$, the first column and the last row has been
deleted. For $N=5$, a column $(+1,-1,-1,-1)$ has been adjoined.
The sets for $N=7$ and $N=9$ are obtained from the rows of the Sylvester type
Hadamard matrix $H_{2^3}$ -- confer eq.~(\ref{eq:H8}). For $N=7$, the 8th column and row have been
deleted, and for $N=9$, a column with alternating $\pm1$'s has been adjoined.

The largest shattered sets we could find for the balanced Ising percpetron with
binary patterns are:

\begin{tabular}{ccc}
$N=9$ & $N=15$ & $N=17$ \\
\pat{---------} & \pat{-+-+-+-+-+-+-+-} & \pat{--+-+-+-+-+-+-+-+}\\
\pat{------+++} & \pat{--++--++--++--+} & \pat{+--++--++--++--++}\\
\pat{----++--+} & \pat{-++--++--++--++} & \pat{--++--++--++--++-}\\
\pat{--++----+} & \pat{----++++----+++} & \pat{+----++++----++++}\\
\pat{-+-+-+-+-} & \pat{-+-++-+--+-++-+} & \pat{--+-++-+--+-++-+-}\\
\pat{-+-++-+--} & \pat{--++++----++++-} & \pat{+--++++----++++--}\\
\pat{-++--++--} & \pat{--------+++++++} & \pat{--++-+--+-++-+--+}\\
                & \pat{-+-+-+-++-+-+-+} & \pat{+--------++++++++}\\
                & \pat{--++--++++--++-} & \pat{--+-+-+-++-+-+-+-}\\
                & \pat{-+-++-+-+-+--+-} & \pat{+--++--++++--++--}\\
                & \pat{--++++--++----+} & \pat{--++--++-+--++--+}\\
                &                       & \pat{+----++++++++----}\\
                &                       & \pat{--+-++-+-+-+--+-+}
\end{tabular}

These pattern sets lead to the lower bounds in eq.~(\ref{eq:low_bounds_balanced}).
The set for $N=9$ has been found by exhaustive enumeration and has no simple relation
to a Hadamard matrix. The patterns for $N=15$ are rows 2-11, 14 and 15 of the Sylvester
type Hadamard matrix $H_{2^4}$ with the last column deleted. The patterns for $N=17$ are
rows 2-14 of $H_{2^4}$, extended by a column of alternating $\pm1$'s.

Pattern sets that exceed the bounds given in eq.~(\ref{eq:low_bounds_balanced}) can
be constructed for these values of $N$:
\begin{description}
\item[$N=15$:]  Delete the last column from the Sylvester type Hadamard matrix 
$H_{2^4}$; then the first 13 rows form a shattered pattern set.
\item[$N=23$:] Delete the last column from the Hadamard matrix $H_2\otimes H_{11+1}$;
then the first 17 rows form a shattered pattern set.
\item[$N=27$:] Delete the last column from the Paley type Hadamard matrix $H_{2(13+1)}$;
then the first 19 rows form a shattered pattern set.
\item[$N=31$:] Delete the last column from the Sylvester type Hadamard matrix 
$H_{2^5}$; then the rows number 2 to number 25 form a shattered pattern set with 24
patterns.
\end{description}

\bibliographystyle{unsrt}
\bibliography{bib/vc,bib/spinglass,bib/nn,bib/math}

\end{document}